\documentclass[sigconf]{acmart}

\AtBeginDocument{%
  \providecommand\BibTeX{{%
    \normalfont B\kern-0.5em{\scshape i\kern-0.25em b}\kern-0.8em\TeX}}}

\usepackage {siunitx} 

\copyrightyear{2023}
\acmYear{2023}
\acmDOI{10.1145/3581961.3609887}
\setcopyright{rightsretained}
\acmConference[AutomotiveUI '23 Adjunct]{15th International Conference on Automotive User Interfaces and Interactive Vehicular Applications}{September 18--22, 2023}{Ingolstadt, Germany}
\acmBooktitle{15th International Conference on Automotive User Interfaces and Interactive Vehicular Applications (AutomotiveUI '23 Adjunct), September 18--22, 2023, Ingolstadt, Germany}\acmDOI{10.1145/3581961.3609887}
\acmISBN{979-8-4007-0112-2/23/09}

\begin{document}

\title[Neuroergonomic Evaluation of the Speed Regulation N-Back Task]{Manipulating Drivers' Mental Workload: Neuroergonomic Evaluation of the Speed Regulation N-Back Task Using NASA-TLX and Auditory P3a}


\author{Nikol Figalová}
\authornote{Both authors contributed equally to this research.}
\email{nikol.figalova@uni-ulm.de}
\orcid{0000-0001-7618-4852}
\affiliation{%
  \institution{Clinical and Health Psychology, Inst. of Psychology and Education, Ulm University}
  \streetaddress{Albert-Einstein-Allee 45}
  \city{Ulm}
  \country{Germany}}

\author{Jürgen Pichen}
\authornotemark[1]
\email{juergen.pichen@uni-ulm.de}
\orcid{0000-0003-2844-0465}
\affiliation{%
  \institution{Human Factors, Inst. of Psychology and Education, Ulm University}
  \streetaddress{Albert-Einstein-Allee 45}
  \city{Ulm}
  \country{Germany}
  \postcode{89081}
}

\author{Vanchha Chandrayan}
\orcid{0000-0003-4242-5747}
\affiliation{%
  \institution{Human Factors, Inst. of Psychology and Education, Ulm University}
  \streetaddress{Albert-Einstein-Allee 45}
  \city{Ulm}
  \country{Germany}}

\author{Olga Pollatos}
\orcid{0000-0002-0512-565X}
\affiliation{%
  \institution{Clinical and Health Psychology, Inst. of Psychology and Education, Ulm University}
  \streetaddress{Albert-Einstein-Allee 45}
  \city{Ulm}
  \country{Germany}}

\author{Lewis Chuang}
\orcid{0000-0002-1975-5716}
\affiliation{%
 \institution{Human and Technology, Inst. for Media Research, Faculty of Humanities, Chemnitz University of Technology}
 \city{Chemnitz}
 \country{Germany}}

\author{Martin Baumann}
\orcid{0000-0002-2668-2527}
\affiliation{%
 \institution{Human Factors, Inst. of Psychology and Education, Ulm University}
  \streetaddress{Albert-Einstein-Allee 45}
  \city{Ulm}
  \country{Germany}}

\renewcommand{\shortauthors}{Figalová and Pichen, et al.}

\begin{abstract}
Manipulating MW in driving simulator studies without the need to introduce a non-driving-related task remains challenging. This study aims to empirically evaluate the modified speed regulation n-back task, a tool to manipulate drivers' MW. Our experiment involved 23 participants who experienced a 0-back and 2-back driving condition, with task-irrelevant novel environmental sounds used to elicit P3a event-related potentials. Results indicate that the 2-back condition was perceived as more demanding, evidenced by higher NASA-TLX scores (overall score, mental and temporal demand, effort, frustration). The mean P3a amplitude was diminished during the 2-back condition compared to the 0-back condition, suggesting that drivers experienced higher MW and had fewer resources available to process the novel environmental sounds. This study provides empirical evidence indicating that the speed regulation n-back task could be a valid, effective, and reproducible method to manipulate MW in driving research.
\end{abstract}




\begin{teaserfigure}
  \includegraphics[width=\textwidth]{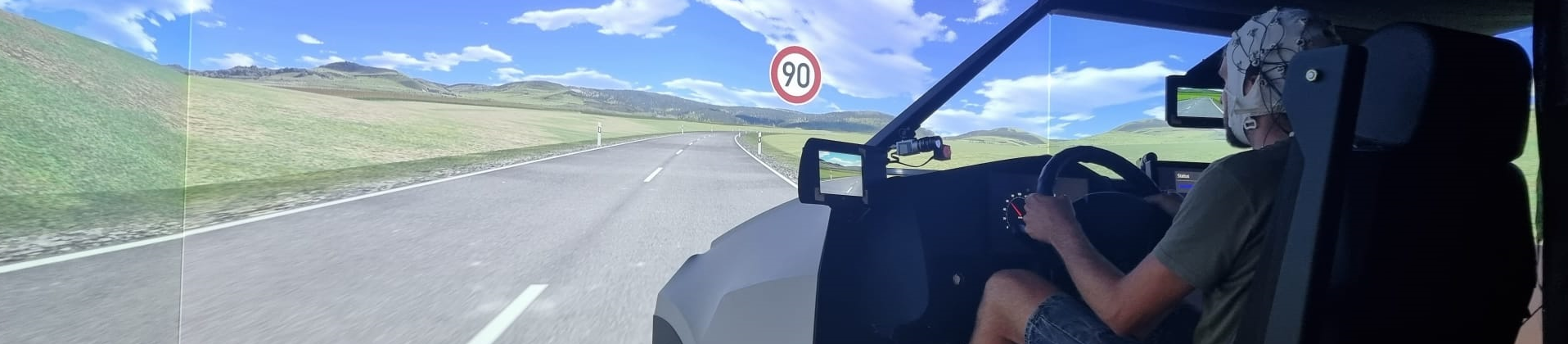}
  \caption{The experimental setup: A participant is conducting the modified speed regulation n-back task in a driving simulator while his electroencephalography (EEG) is being recorded.}
  \Description{}
  \label{fig:teaser}
\end{teaserfigure}

\received{}
\received[revised]{}
\received[accepted]{}

\maketitle

\section{Introduction}
There are various methods to manipulate mental workload (MW) in the field of driving research. Traditionally, researchers would ask participants to perform a non-driving related task (NDRT) while driving or monitoring an automated vehicle to manipulate the level of experienced MW. This approach can be described as a dual-task approach. Contrary, a single-task approach requires drivers to be involved only in one task, typically driving. In order to increase or decrease the demands of driving, researchers would manipulate environmental factors, such as traffic complexity \cite{teh2014temporal}. 

Implementing MW manipulation within a single-task approach presents significant challenges. Researchers often modify the difficulty of the task based on their best understanding (for instance, by increasing traffic in the simulated scenario or reducing visibility) and presume the manipulation yields the intended effect. However, these manipulations are not standardized, making it problematic to evaluate or replicate the effects. Conversely, applying a standardized secondary NDRT to manipulate MW addresses issues of reproducibility and subjectivity in experimental designs. Yet, accurately estimating the distribution of participants' mental resources between primary and secondary tasks can be problematic. Moreover, it potentially undermines the realism of the driving experience \cite{hock2018design}.

Acquiring controlled, standardized MW manipulation within a single-task driving paradigm remains challenging. This paper aims to provide further empirical evidence for the impact of a single-task MW manipulation using a modified speed regulation n-back driving task proposed by Unni et al. \cite{unni2017assessing}. In the present study, participants  encountered a 0-back and 2-back condition within a driving simulator. We analysed self-reported MW (using NASA-TLX \cite{hart2006nasa}) and objective MW (measured via P3a amplitude induced by task-irrelevant environmental sounds \cite{polich2007updating, scheer2016steering}). To our best knowledge, this study is the first attempt to assess the speed regulation n-back task employing P3a event-related potential (ERP). Moreover, no self-reported assessment of perceived MW induced by the n-back task has been previously published.

Understanding the effect of MW manipulation via the speed regulation n-back task on the perceived cognitive load and P3a amplitude will provide valuable insights into the cognitive mechanisms underlying this task. Providing empirical evidence combining objective and subjective measures of MW will help to establish the speed-regulation n-back task as an objective, replicable way of MW manipulation in driving simulator studies. 

\subsection{Related Work} \label{relatedwork}
Mental workload (MW) refers to the amount of cognitive resources required to perform a specific task \cite{embrey2006development}. It's a measure of the mental effort expended, taking into account factors such as task complexity, information processing demands, and the cognitive capabilities of the individual. MW is measurable through various techniques, including self-reporting, performance-based assessments, and physiological measurements \cite{de1996measurement}. Using event-related potentials (ERPs) evoked by novel environmental auditory cues has been proposed to be an objective physiological measure of MW \cite{polich2007updating, escera1998neural, kramer1995assessment}. The authors of these studies jointly report that higher MW correlates with a lower amplitude of the P3a ERP component and vice versa. This effect is based on the multiple resources theory \cite{wickens2008multiple}, which suggests that the environmental stimuli compete for limited processing resources. When a high mental workload is induced, fewer attentional resources are left for processing the environmental sounds \cite{chun2011taxonomy}. Higher MW, therefore, leads to decreased P3a amplitude. There is a discussion about whether other factors, such as arousal or top-down attention control, might be a factor influencing the P3a amplitude \cite{cahn2009meditation, figalova2023driver}; nevertheless, robust empirical evidence supports the inversed relationship between MW and P3a amplitude due to the competition for resources.

Several authors studied the P3a amplitude in the driving research. \citeauthor{wester2008event} compared the P3a amplitudes between stationary and driving conditions. \citeauthor{scheer2016steering} studied the effect of steering demands on P3a amplitude. \citeauthor{van2018susceptibility} compared the P3a amplitude in a stationary condition to manual and automated driving. \citeauthor{figalova2023driver} studied the effect of different levels of automation on the P3a amplitude. However, none of these studies used a standardised single-task method to manipulate MW. 

An n-back task is a continuous performance task designed to measure a part of working memory and working memory capacity. The load factor \textit{n} can be adjusted to make the task more or less difficult and, therefore, to manipulate MW. In the modified speed regulation n-back driving task proposed by Unni et al. \cite{unni2017assessing}, participants manually drive while presented with a series of speed signs. The task requires participants to continuously update, memorise, and recall previous speed signs while adjusting their current speed to match the sign displayed \textit{n}-steps before. This paradigm has been used in other studies \cite{unni2017assessing, held2022multitasking, scheunemann2019demonstrating} and appears to be an efficient way to manipulate MW using a single-task approach.

\subsection{Hypotheses}
Participants experienced a low MW condition (0-back) and a high MW condition (2-back). Based on the inversed relationship between MW and P3a amplitude discussed in section \ref{relatedwork}, we formulated the following hypotheses: 
\begin{itemize}
    \item \textbf{H1:} The overall score of NASA-TLX is lower during the 0-back than during the 2-back condition.
    \item \textbf{H2:} The mean P3a amplitude is higher during the 0-back than during the 2-back condition.
\end{itemize}

\section{Methods}

\subsection{Particpants}
Our sample consisted of 23 participants (11 males and 12 females) with an average age of 23.83 years (\textit{SD} = 3.61). All participants had normal or corrected-to-normal vision, no known neurological or psychiatric disease, and a valid German driving license (on average for \textit{M} = 3.77 years, \textit{SD} = 3.48). Participants provided informed consent and received 50 Euros as compensation to participate.

\subsection{Apparatus and Procedure}
The experiment took place in a highly immersive, fixed-base driving simulator with a realistic car mock-up (see Figure \ref{fig:teaser}). The EEG was recorded using 32 active channels (ActiCAP by Brain Products) placed according to the international 10-20 system. We kept the impedance below $25 k\Omega$. We also recorded electrodermal activity and electrocardiogram; however, the data are not presented here. The perceived cognitive workload was assessed using the NASA Task Load Index scale (NASA-TLX; \cite{hart2006nasa}). Each of the six dimensions was evaluated on a 100-point Likert scale. The total score was obtained as the average of the six dimensions. 

First, participants learned how to operate the driving simulator and conduct the speed regulation n-back task. Afterwards, we installed the EEG and started the experiment. Participants filled in the NASA-TLX at the end of each condition. The order of the conditions was counterbalanced between participants.

\subsection{Experimental task}
\begin{figure}
    \centering
  \includegraphics[width=1 \linewidth]{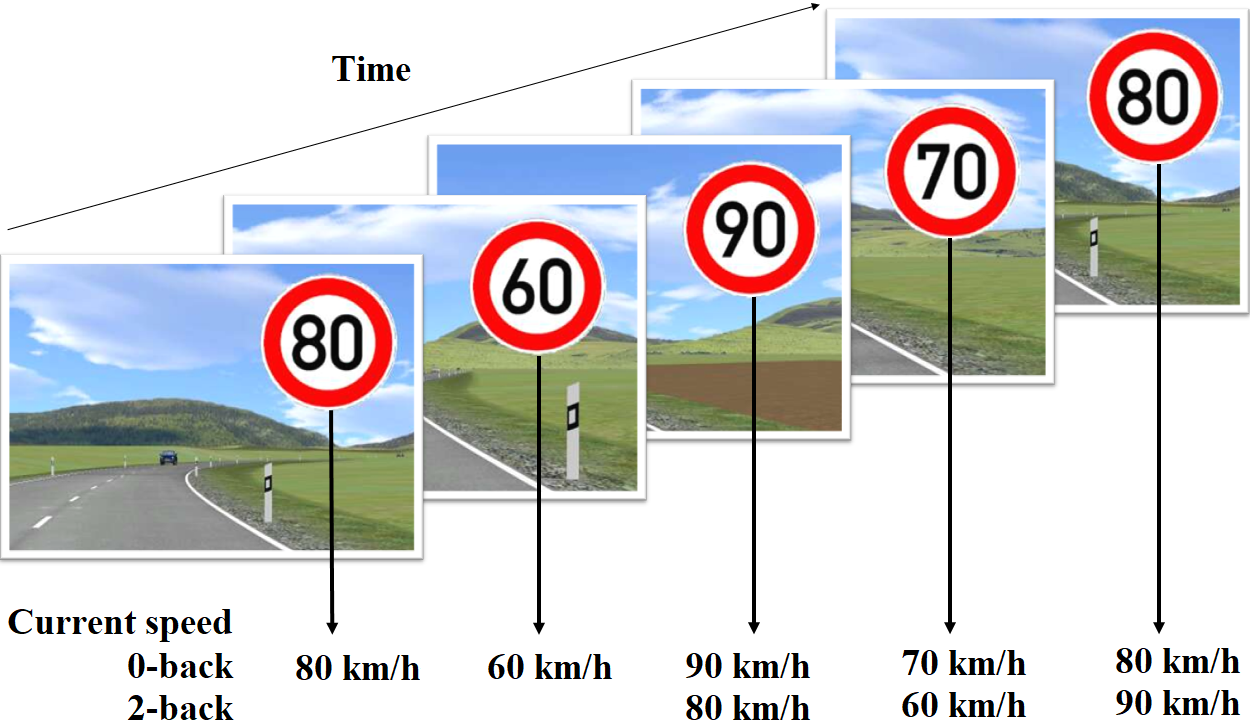}
  \caption{Example of the speed regulation n-back task.}
  \Description{}
  \label{NBackfigure}
\end{figure}

Participants experienced speed regulation 0-back and 2-back conditions designed to manipulate MW using a within-subject design. Participants drove on a rural road, and a speed sign (from 60 to 100 km/h) was presented in a random order every 20 seconds on a head-up-like screen in front of them (see Figure \ref{fig:teaser}). The position and size of the sign corresponded to a normal German speed sign. The sign was shown for 3 seconds. For the 0-back task, participants were instructed to adjust and maintain the speed according to the currently presented sign. For the 2-back task, participants had to adjust and maintain the speed that occurred two signs before and remember the sequence of the two following speed signs. Each condition lasted for approximately 17 minutes, and 52 signs were shown in each condition. The task is visualised in Figure \ref{NBackfigure}.

\subsection{Stimuli and Questionnaires}
Participants were exposed to a passive oddball task, which consisted of three types of task-irrelevant auditory stimuli:

\begin{enumerate}
\item[(1)]\textbf{Frequent distractors,} 450 presentations in each condition (probability of presentation: 71.43\%);
\item[(2)]\textbf{infrequent distractors,} 90 presentations in each condition (probability of presentation: 14.29\%);
\item[(3)]\textbf{novel environmental distractors,} 90 presentations in each condition (probability of presentation: 14.29\%).
\end{enumerate}

The frequent and infrequent distractors consisted of two tones (pure 700 Hz and 300 Hz tones); their presentation probability was counterbalanced across participants. The novel environmental distractors consisted of 30 complex sounds (e.g., human laughter, helicopter) \cite{fabiani1996naming}. Each stimulus was presented for 336 ms. The inter-stimulus interval was randomized and ranged from 1300 to 1700 ms. 

\subsection{EEG Pre-processing and Analysis}
Data preprocessing followed the BeMoBil pipeline \cite{klug2022bemobil}. The data were submitted for adaptive mixture independent component analysis (AMICA). We classified the components using IClabel \cite{pion2019iclabel} and retained the components that most likely originated from brain activity. Next, we filtered the data between 0.1 and 30 Hz and extracted epochs from -200 to 800 ms relative to the stimulus onset. Artifactual epochs were rejected. The preprocessed data were averaged for each channel and experimental condition. A difference wave was computed as the difference between the ERP elicited by the environmental distractor and the ERP elicited by the frequent distractor. Statistical analysis was conducted focusing on the Fz electrode using JASP 0.16.4, with the general level of significance set to 0.05.

\section{Results}
\subsection{Self-reported Mental Workload}
Using a paired samples t-test, we found a difference in the \textbf{total NASA-TLX score} between 0-back (\textit{M} = 42.464, \textit{SD} = 15.187) and 2-back (\textit{M} = 58.478, \textit{SD} = 14.466) experimental conditions; \textit{t}(22) = 7.326, \textit{p} \textless .001, \textit{d} = 1.528. Comparing the six dimensions of NASA-TLX, we found a difference in \textbf{mental demand} (\textit{M} = 36.957, \textit{SD} = 21.253 for 0-back; \textit{M} = 77.826, \textit{SD} = 19.588 for 2-back; \textit{t}(22) = 9.700, \textit{p} \textless .001, \textit{d} = 2.023), \textbf{temporal demand} (\textit{M} = 39.348, \textit{SD} = 21.443 for 0-back; \textit{M} = 53.478, \textit{SD} = 25.603 for 2-back; \textit{t}(22) = 3.472, \textit{p} = .002, \textit{d} = 0.724), \textbf{effort} (\textit{M} = 45.652, \textit{SD} = 24.416 for 0-back; \textit{M} = 73.261, \textit{SD} = 20.204 for 2-back; \textit{t}(22) = 6.260, \textit{p} \textless .001, \textit{d} = 1.305), and \textbf{frustration} (\textit{M} = 29.565, \textit{SD} = 31.315 for 0-back; \textit{M} = 47.391, \textit{SD} = 27.214 for 2-back; \textit{t}(22) = 3.790, \textit{p} \textless .001, \textit{d} = 0.790). The results are visualised in \ref{ERPfigure}a. 

The results suggest that the overall perceived mental workload was lower in the 0-back task than in the 2-back task. Moreover, this finding is consistent with the differences observed in the single factors of NASA-TLX. Therefore, we\textbf{ accept H1.}

\subsection{Mean P3a amplitude}
\begin{figure}
    \centering
  \includegraphics[width=1 \linewidth]{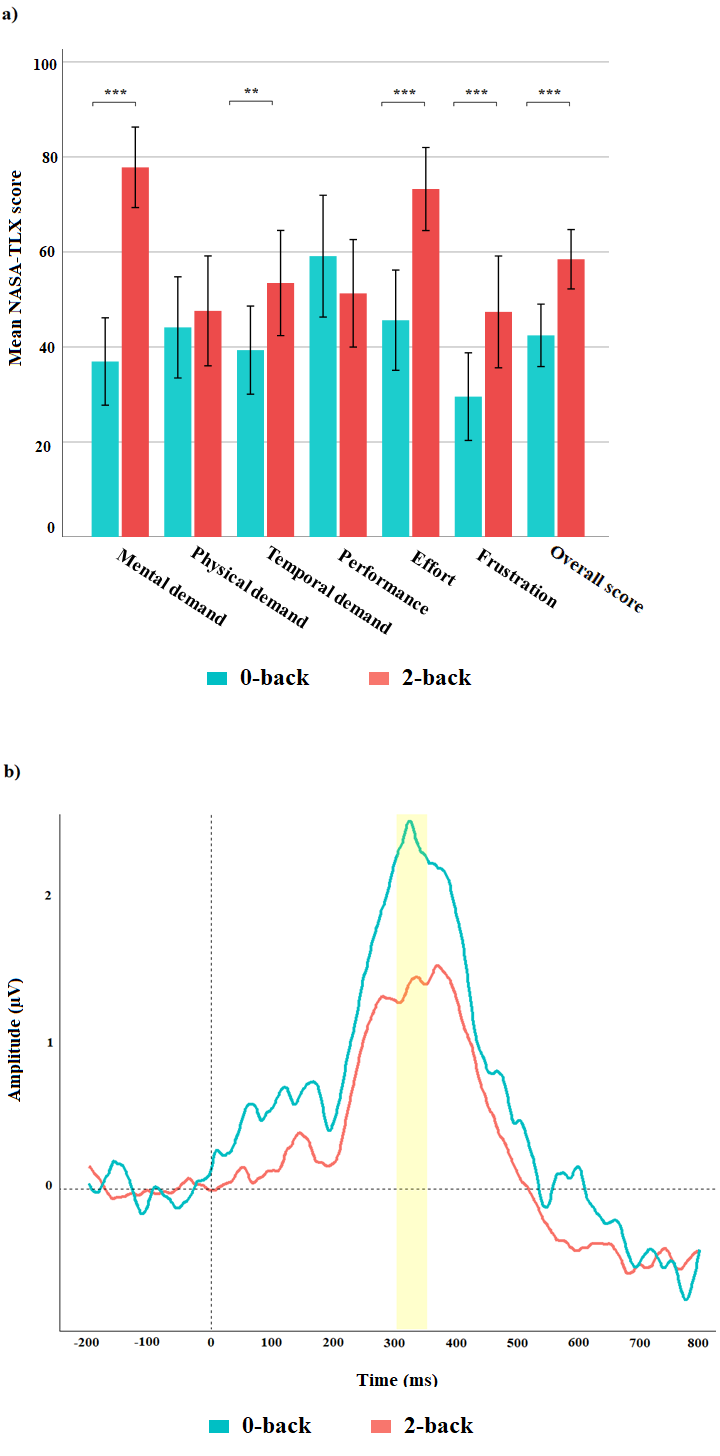}
  \caption{\textit{a)} The mean score of the six dimensions of NASA-TLX. The error bars represent a 95\% confidence interval. Significant differences are highlighted (*** \textit{p} $\leq$ .001, ** \textit{p} $\leq$ .01). \\ \textit{b)} Difference wave (environmental novel distractor-frequent distractor) ERPs for the 0-back and 2-back experimental conditions at the Fz electrode. Mean amplitudes in the yellow-highlighted area (303 to 353 ms) were submitted for statistical comparisons.}
  \Description{}
  \label{ERPfigure}
\end{figure}

We used the collapsed localizer method to quantify ERP component amplitudes. A grand average peak (GA) was determined at 328 ms post-stimulus. The mean amplitude was calculated between 303 and 353 ms post-stimulus (GA peak +/- 25 ms). Figure \ref{ERPfigure}b presents the difference waves (environmental-frequent). We found a difference in the mean amplitude between 0-back (\textit{M} = 1.429 \si{\micro\volt}, \textit{SD} = 1.272) and 2-back (\textit{M} = 2.471 \si{\micro\volt}, \textit{SD} = 2.471) experimental conditions; \textit{t}(22) = 1.937, \textit{p} = .033, \textit{d} = 0.404. 

The results suggest that the P3a amplitude elicited by the novel environmental sounds was decreased with increased mental workload. Therefore, we \textbf{accept H2}. 

\section{Discussion}
This study investigated the effect of MW manipulation using the modified speed regulation n-back task proposed by Unni et al. \cite{unni2017assessing}. We conducted a driving simulator experiment with 23 participants who experienced 0-back and 2-back conditions. We analysed the self-reported MW (using NASA-TLX \cite{hart2006nasa}). Moreover, we studied the mean amplitude of P3a ERP components evoked by task-irrelevant novel sounds, which served as a proxy measure of objective MW \cite{kramer1995assessment,scheer2016steering}. 

The subjective assessment of MW using the overall NASA-TLX suggests that drivers perceived the 2-back condition as more demanding than the 0-back condition. The mental and temporal demands were higher for the 2-back condition. Moreover, participants felt more effort was required to complete the 2-back condition and were more frustrated by the 2-back condition compared to the 0-back condition. These results were anticipated. However, none of the previous studies employing the speed regulation n-back task \cite{unni2017assessing, held2022multitasking, scheunemann2019demonstrating} reported scores of NASA-TLX or other comparable measures. 

Moreover, this study was the first attempt to use an objective measure of MW, namely the mean P3a amplitude evoked by novel environmental sounds, to assess the speed regulation n-back task. The P3a results suggest that the different levels of the n-back task had a significant effect on the amount of resources used to process the novel environmental sounds. Employing the multiple resources theory \cite{wickens2008multiple}, we argue that the 2-back condition required more processing resources; therefore, fewer resources were left to process the novel environmental sound. This effect can be observed as a decreased mean P3a amplitude in the 2-back condition compared to the 0-back condition. This effect was anticipated based on the literature review. 

The differences in subjective MW between the two conditions suggest a large effect (\textit{d} = 1.528) for the overall score of NASA-TLX. Hence, we argue that there were substantial differences in demands between the two conditions and, therefore, the changes in the P3a amplitude can be explained in the light of the multiple resources theory \cite{wickens2008multiple} and resource competition \cite{chun2011taxonomy}. However, we suggest that in the case of low-demanding tasks, researchers could potentially observe a floor effect. This could be due to the low demands of the task, which does not fully occupy the processing resources. As plenty of free resources are available to process the auditory cues, the P3a amplitude might not be in a direct reversed relationship with the demands of the primary task, and could therefore be non-sensitive to the demands of the primary task. 

We observed an inversed relationship between the task demands and the P3a amplitude. This finding is in line with the majority of other studies (e.g., \cite{kramer1995assessment, polich2007updating, escera1998neural}. However, some authors argue that other factors might be modulating the P3a amplitude as well. The results of \citeauthor{cahn2009meditation} suggest that attentional withdrawal during meditation leads to a decreased P3a amplitude evoked by novel auditory stimuli. The results of \citeauthor{figalova2023driver} suggest a discrepancy between perceived MW and mean P3a amplitude evoked by novel auditory stimuli during automated driving in a realistic environment. These studies suggest that other factors, namely top-down attention control, novelty, environment, or arousal, might play an important role in modulating the P3a amplitude. These factors should be addressed methodologically in future studies. 

The empirical data support both H1 and H2 and provide further empirical evidence for the reliability and validity of the speed regulation n-back task to manipulate MW in driving research. Therefore, the speed-regulation n-back task may be used to manipulate MW in a standardised way. Nevertheless, further methodological recommendations must be derived in order to ensure the results of different studies conducted in different environments utilising different scenarios are directly comparable.

\subsection{Limitations and Future Work}
This study focused only on two levels of the n-back task (0-back and 2-back). For a fine-grained evaluation of the task, we recommend measuring more levels (e.g., 0, 1, 2, and 3 with each participant). Moreover, we recommend validating the paradigm on a general adult sample, as our sample consisted predominantly of young university students. Future studies should also address the effect of other variables, such as arousal or top-down attention control, on the auditory P3a amplitude.

\subsection{Contribution and Novelty}
This study evaluates the speed regulation n-back task, a novel approach to manipulating MW in driving research. We used methods of estimating MW induced by the speed regulation n-back task that have not been reported before in the context of the task (NASA-TLX and mean P3a amplitude). We combined self-reported and EEG-based estimates of mental workload that provide robust evidence of the validity and reliability of the task. Moreover, our empirical data help to comprehensively understand the cognitive mechanisms underlying the task. 

\section*{Conclusion}
The data suggest that drivers perceived the 2-back condition as more demanding than the 0-back condition. We found differences in the mental and temporal demand, effort, and frustration dimensions of NASA-TLX. Moreover, the P3a amplitude was significantly decreased for the 2-back condition compared to the 0-back condition. The speed regulation n-back task appears to be a reliable, valid, and reproducible method to manipulate MW in driving research. 

\begin{acks}
  This project has received funding from the European Union's Horizon 2020 research and innovation programme under the Marie Sklodowska-Curie grant agreement 860410. The project was supported by the Deutsche Forschungsgemeinschaft (DFG, German Research Foundation)-Project-ID 416228727-SFB 1410.
\end{acks}

\bibliographystyle{ACM-Reference-Format}
\bibliography{references}

\end{document}